\documentclass[showpacs,floatfix,aps,prl,preprint]{revtex4}
\begin{document}
%\preprint{hep-ph/0210160}
\title{Evolving Fundamental Constants and Metrology}
\author{A.~Yu.~Ignatiev}
\email{a.ignatiev@physics.unimelb.edu.au}
\author{B.~J.~Carson}
\email{b.carson@physics.unimelb.edu.au} \affiliation{\em School of
Physics, Research Centre for High Energy Physics, The University of
Melbourne, Victoria 3010, Australia.}
%\date{\today}
% \renewcommand{\today}{}
% \setlength\textwidth{6.5 in}
% \setlength\topmargin{-0.5cm}
% \setlength\textheight{9 in}
% \addtolength\evensidemargin{-1.cm}
% \addtolength\oddsidemargin{-1.cm}
% \font\tenrm=cmr10
% \setlength{\parindent}{0pt}
% \setlength{\parskip}{6pt}
%\usepackage{epsfig}
\def\be{\begin{equation}}
\def\ee{\end{equation}}
\def\bea{\begin{eqnarray}}
\def\eea{\end{eqnarray}}
\newcommand{\nn}{\nonumber \\}
\begin{abstract}

Astrophysical observations suggest that the fine structure constant
(alpha) may (or may not) be evolving over the cosmological timescale.
This raises a much debated question: is alpha variation due to the
variation of the speed of light (c), elementary electric charge (e), or
the Planck constant (h)?

Previously, we proposed the metrological approach based on the analysis
of the relationships between the fundamental units (e.g. of the length
and time) and the fundamental constants. Our methodology allows one to
find how each of the fundamental constants e, c, h evolves in time and
offers a new outlook for this area. Here we give a brief outline of this
approach and the main results it produces.
\end{abstract} \maketitle
%\documentclass[prl,twocolumn]{revtex4}
%\tighten
%\preprint{UM-P, RCHEP-}
%\doublespace
%\date{}

Recent astrophysicsl observations suggest that the fine structure
constant $\alpha$ may be evolving over the cosmological timescale.
According to the results of one series of studies  of absorbtion spectra
of distant quasars, $\Delta\alpha/\alpha=(-0.57\pm 0.11)\times 10^{-5}$
over the redshift range $0.2<z<4.2$ \cite{webb1,webb2,mur}. Another study
finds that $\alpha$ does not change, up to the observational accuracy:
$\Delta\alpha/\alpha = (-0.06 \pm 0.06)\times 10^{-5}$ over the redshift
range $0.4<z<2.3$ \cite{sri}. The variation of the fine structure
constant in $^{187}Re$ decay over the age of the solar system
(corresponding to $0<z\alt 0.4$) has been found consistent with zero at
the level of $\Delta\alpha/\alpha\sim 10^{-7}$ \cite{oli}. There are also
studies of the possible variations of other dimensionless constants such
as the proton-to-electron mass ratio $\mu=m_e/m_p$ \cite{iva}.

For other theoretical and experimental aspects of the problem, see
\cite{uzan,mag,we,tob} and references therein \footnote{Mota and Barrow
    recently proposed the important idea of considering the
effect of inhomogeneity on the evolution of the fundamental constants
\cite{mb1,mb2,mb3}.}.

Since $\alpha=e^2/\hbar c$ the idea of variable $\alpha$ raises a much
debated \cite{mag,duff1,duff2,mof,nat,fla} question: is $\alpha$
variation due to the variation the speed of light (c), elementary
electric charge (e) or the Planck constant ($\hbar$)?

 The metrological approach \cite{ic} is based on 2 self-explanatory postulates:

 1. Before (or in parallel to) developing a theory of variable constants
 one needs to specify how the time, length and other physical quantities
 are measured. Otherwise the analysis of constants evolution can be
 ambiguous or impossible. This is because the fundamental constants (e,
 c, $\hbar$ \dots )enter the definitions of the basic units (metre,
 second, \dots ). Therefore, both sets should be considered as one system
 within a self-consistent analysis.

 2. The basic units should be considered time-independent. Otherwise it
 would be hard to give precise meaning to the time evolution of the
 dimensional constants such as e, c, $\hbar$.

For simplicity and correspondence with the previous literature, the
centimetre-gram-second (CGS) system of units will be used. (This
assumption, however, is not important for the essence of our argument.)
Also, as there is no strong observational or experimental evidence for
the time-dependence of the fundamental constants apart from $\alpha$,
such as the mass ratios ($m_e/m_p$, $m_e/m_{Nucl}$, \dots) and
$g$-factors it will be assumed here that they do not vary with time (for
theoretical discussions see,e.g., \cite{cal,lan,dent}).
% (alternatives will be discussed at the end of the paper):
% [add Okun's argument in favor of 3 unit system];
% \footnote{$\alpha$ sits in g -
%neglect as higher order effect?}

Our units of time and length can be defined in several different ways
depending on what physical processes (or objects) are chosen to represent
the standards of time and length measurement.  Consequently, the
dependence of the second and centimetre on the fundamental constants will
take different forms. It can be shown that the three practical
definitions of the second are described by \be\label{n} s_n \propto
\frac{\hbar}{m_e\alpha^n c^2}, \ee where $n=1$ corresponds to the
Superconducting Cavity Stabilised Oscillator (SCSO) clock; $n=2$, to
ammonia clock ($NH_4$ molecular vibration) and $n=3$, to
 cesium clock (cesium-133 hyperfine transitions) \cite{fn}.

Similarly, the three different definitions of the centimetre can be
combined as \be\label{l} \underline{cm}_l \propto
\frac{\hbar}{m_e\alpha^l c} \ee with $l=1$ describing the ``bar'' ruler
(i.e., the 1/100 of the length between two notches on the
platinum-iridium metre prototype), $l=2$ --- the ``krypton'' ruler (the
length equal to the 16507.6373 wavelengths
  of the radiation corresponding to the transition between the levels
   $2p_{10}$ and $5d_5$ of krypton-86), and $l=3$ --- the
``light'' ruler (the length travelled by light in a vacuum during a time
interval of 1/299 792 45800 of a cesium second).

From dimensional considerations it follows that an arbitrary system of
units, $S_{nl}$, can be characterized by Eq.(\ref{n}) and (\ref{l}) with
a suitable choice of clock and ruler indices $n$ and $l$.

According to the ``second metrological postulate'' we should require that
the centimetre, second and gram do not depend on time, and obtain \bea  c
& \propto \alpha^{l-n} \label{main}\\ \hbar & \propto \alpha^{2l-n}\\
 e & \propto \alpha^{\frac{1+3l-2n}{2}}. \label{main1}\eea

In other words, \bea\label{g}\frac{\Delta c}{c}&\simeq (l-n)\frac{\Delta
\alpha}{\alpha} \\ \frac{\Delta \hbar}{\hbar}& \simeq (2l-n)\frac{\Delta
\alpha}{\alpha} \\  \frac{\Delta e}{e}& \simeq
 \left(\frac{1+3l-2n}{2}\right)\frac{\Delta
\alpha}{\alpha}.\eea

Thus, if time dependence of $\alpha$ is experimentally measured {\em and}
units are fixed, there is {\em no} further choice on how $e, c, \hbar$
depend on time and no extra measurements are required.

 In
summary, we propose a metrological approach based on the analysis of the
relationships between  the fundamental units (e.g. of the length and
time) and the fundamental constants. Our methodology allows one to find
how each of the fundamental constants e, c, h evolves in time and offers
a new outlook for this area.
 We are grateful to G.C.Joshi, W.McBride, B.H.J.McKellar and
R.R.Volkas for stimulating discussions and to T.Dent and D.F.Mota for
valuable comments.


\begin{thebibliography}
\bibitem{fn}These equations are
valid neglecting higher order corrections. Inclusion of these corrections
will lead to effective non-integer values of n:$n\rightarrow
n_{eff}=n+\delta_n$.
\end{thebibliography}
\end{document}